# The curvature-induced magnetization in CrI3 bilayer: flexomagnetic effect enhancement in van der Waals antiferromagnets

Lei Qiao[1], Jan Sladek[2], Vladimir Sladek[2], Alexey S. Kaminskiy[3], Alexander P. Pyatakov[3,*], Wei Ren[1,4,*]

[1] *Physics Department, International Center of Quantum and Molecular Structures, Materials Genome Institute, State Key Laboratory of Advanced Special Steel, Shanghai Key Laboratory of High Temperature Superconductors, Shanghai University, Shanghai 200444, China*
[2] *Institute of Construction and Architecture, Slovak Academy of Sciences, 84503 Bratislava, Slovakia, e-mail: jan.sladek@savba.sk*
[3] *MIREA - Russian Technological University, 119454, Moscow, Russia*
[4] *Zhejiang Laboratory, Hangzhou 311100, China*

*Pyatakov@physics.msu.ru; *renwei@shu.edu.cn

The bilayer of $CrI_3$ is a prototypical van der Waals 2D antiferromagnetic material with magnetoelectric effect. It is not generally known, however, that for symmetry reasons the flexomagnetic effect, *i.e.*, the strain gradient-induced magnetization, is also possible in this material. In the present paper, based on the first principle calculations, we estimate the flexomagnetic effect to be 200 $\mu_B\cdot$Å that is two orders of magnitude higher than it was predicted for the referent antiperovskite flexomagnetic material $Mn_3GaN$. The two major factors of flexomagnetic effect enhancement related to the peculiarities of antiferromagnetic structure of van der Waals magnets is revealed: the strain-dependent ferromagnetic coupling in each layer and large interlayer distance separating antiferromagnetically coupled ions. Since 2D systems are naturally prone to mechanical deformation, the emerging field of flexomagnetism is of special interest for application in spintronics of van der Waals materials and straintronics in particular.

*Introduction*--Since the first report on graphene isolation, the class of 2 dimensional (2D) materials has expanded tremendously: it includes now not only graphene derivatives (like graphane, graphone, graphyne, *etc.*) but also other types of van der Waals materials including monolayers and bilayers of transition metal dichalcogenides [1] and dihalides [2]. Some of these compounds have been recently discovered to demonstrate 2D magnetic ordering [3,4]. Since 2D materials are naturally prone to mechanical deformation the study of cross-correlation effects between lattice, electronic and magnetic subsystems is critical for the practical applications in straintronics, an emergent branch of electronics related to the strain-induced effects [5]. The advent of 2D magnets can bridge the gap between the two concepts of straintronics: the straintronics of magnets [6] and the straintronics of van der Waals materials [7,8].

The flexural deformation (bending) characterized by strain gradient induces the electric polarization in the crystal, by the effect known as the flexoelectric one. In analogy to flexoelectricity the flexomagnetic effect as a strain gradient-induced magnetization was theoretically predicted [9,10] and experimentally found [11,12]. In 2D magnetic materials the flexo-related phenomena have only very recently attracted attention of the researchers: the flexomagnetoelectric coupling in $MoS_2$ [13] and curvature-induced spin cycloid ordering in $CrI_3$ [14,15] were predicted by ab initio calculations, the flexomagnetic phase transition from antiferromagnetic to ferromagnetic order in rippled Heusler membranes was observed [16].

In this paper the $CrI_3$ bilayer is proposed as a material whose symmetry allows flexomagnetic effect, i.e. the magnetization linearly proportional to the strain gradient. Flexomagnetic coefficients are obtained by fitting the DFT simulation results with the analytical solution for a simple problem involving the gradient theory.

*Structure and symmetry*--In a monolayer of $CrI_3$ the Cr atoms form a honeycomb structure (Fig. 1(a)). $CrI_3$ has two stacking styles, AB' stacking ($C_{2h}$ point group) with antiferromagnetic (AFM) interlayer interaction and AB stacking ($S_6$ point group) with ferromagnetic (FM) interlayer interaction. Based on the fact that we study bilayers with AFM interlayer interactions, here we choose AB' stacking bilayer [17]. The layer-dependent magnetic ordering was observed in this material [3]: ferromagnetism in the monolayer, antiferromagnetism in the bilayer, and ferromagnetism in the trilayer again etc. The Cr atoms in every layer of multilayer are ferromagnetically coupled, while the interlayer exchange is an antiferromagnetic one. $CrI_3$ bilayer is an antiferromagnet demonstrating linear magnetoelectric effect [18], in other words, the symmetry of antiferromagnetic order parameter $L$ allows $E_iH_jL_k$ -type invariant combinations, where $E_i$ and $H_j$ are components of electric and magnetic fields respectively. The existence of this invariant implies the magnetization linear with respect to the electric field.

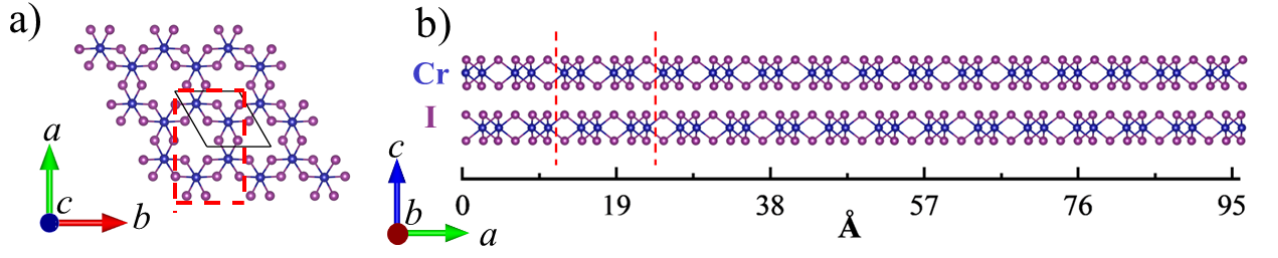

Fig. 1. (a) The top view of CrI3 monolayer. (b) The cross-section of CrI3 bilayer. Blue and purple spheres are Cr and I atoms respectively. The red dashed lines indicate the unit cell.

The strain gradient $\nabla\sigma$ violates the spatial inversion symmetry in a way it does the electric field, so $\nabla\sigma HL$-type invariant is also possible. This invariant allows for the magnetization linear with respect to the strain gradient, *i.e.*, flexomagnetic effect. To investigate the exact structure of the flexomagnetic tensor let us consider the thermodynamic terms related to magnetic and mechanical subsystems in more detail.

*Tensor of flexomagnetic effect and flexomagnetic coefficients*--The flexomagnetism can be phenomenologically described by incorporating additional strain gradient terms into the expression for the thermodynamic potential. Then, the free energy density for a piezomagnetic solid can be written as [10]:

$$F = \frac{1}{2}c_{ijkl}\varepsilon_{ij}\varepsilon_{kl} - \frac{1}{2}\gamma_{ij}H_{i}H_{j} + \frac{1}{2}g_{jklmni}\eta_{jkl}\eta_{mni} - \xi_{ijkl}H_{i}\eta_{jkl} \quad (1)$$

where **H** is a magnetic field, tensor **γ** components are the second-order magnetic permeabilities, **c** stands for the fourth-order elastic tensor, and **ξ** is the flexomagnetic effect tensor. The higher order elastic coefficients corresponding to the strain gradient **η** are denoted by **g**. No piezomagnetic properties are considered.

The linear strain tensor $\varepsilon_{ij}$ is defined as:

$$\varepsilon_{ij} = \frac{1}{2}\left(u_{i,j} + u_{j,i}\right) \quad (2)$$

where $u_i$ is the displacement and the stationary magnetic field is expressed as the negative gradient of the magnetic potential:

$$H_i = -\psi_{,i}$$

The strain-gradient tensor **η** is defined as

$$\eta_{ijk} = \varepsilon_{ij,k} = \frac{1}{2}\left(u_{i,jk} + u_{j,ik}\right) \quad (3)$$

The constitutive equations can be obtained from the free energy density expression (1)

$$\sigma_{ij} = \frac{\partial F}{\partial \varepsilon_{ij}} = c_{ijkl}\varepsilon_{kl}$$

$$\tau_{jkl} = \frac{\partial F}{\partial \eta_{jkl}} = -\xi_{ijkl}H_i + g_{jklmni}\eta_{nmi} \quad (4)$$

$$B_i = -\frac{\partial F}{\partial H_i} = \gamma_{ij}H_j + \xi_{ijkl}\eta_{jkl}$$

where $\sigma_{ij}$, $B_i$, and $\tau_{jkl}$ are the stress tensor, magnetic induction, and higher order stress tensor, respectively.

To simplify the strain gradient theory the internal length material parameter $l$ has been introduced [19,20]. Then, the higher-order elastic coefficients $g_{jklmni}$ can be expressed in terms of the conventional elastic stiffness coefficients $c_{klmn}$ and this material parameter: $g_{jklmni} = l^2 c_{jkmn}\delta_{li}$ with $\delta_{li}$ being the Kronecker delta.

Two independent coefficients $\xi_1$ and $\xi_2$ are introduced for the flexomagnetic tensor $\xi_{ijkl}$:

$$\xi_{ijkl} = \xi_1 \delta_{jk}\delta_{il} + \xi_2 \left(\delta_{ij}\delta_{kl} + \delta_{ik}\delta_{jl}\right) \quad (5)$$

In the framework of this theory the free energy density has the following form

$$F = \frac{1}{2}c_{ijkl}\varepsilon_{ij}\varepsilon_{kl} - \frac{1}{2}\gamma_{ij}H_iH_j + \frac{l^2}{2}c_{jkmn}\eta_{jkl}\eta_{mnl} - \xi_1 H_i\eta_{kki} - 2\xi_2 H_i\eta_{ikk} \quad (6)$$

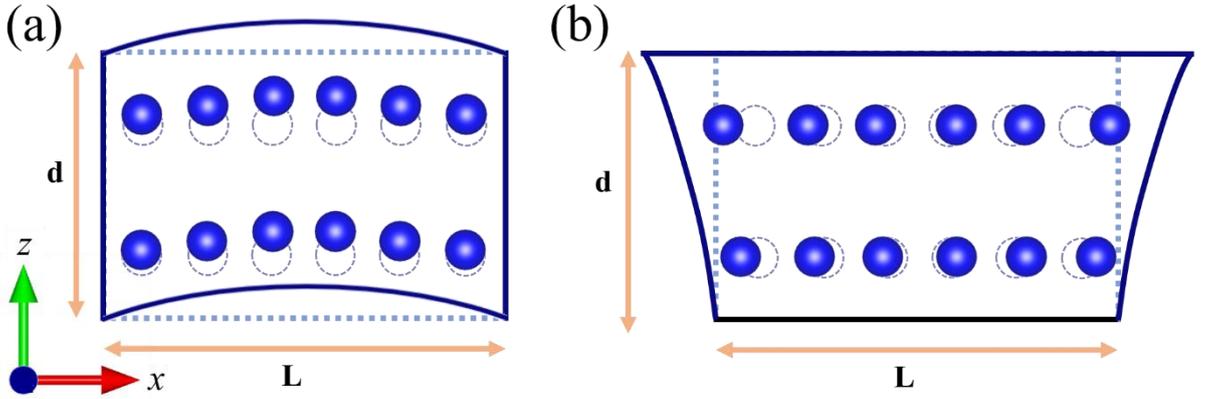

Fig. 2. Schematic diagram of two deformation modes. Symmetric boundary conditions for a) a simple patch problem A: $u_3 = a_1 x^2 + a_2 z^2$, where $a_1 = -(c_{33}/c_{44})a_2$, and b) a simple patch problem B, $u_1 = a_3 x^2 + a_4 z^2$ where $a_3 = -(c_{44}/c_{11})a_4$. Dashed and solid lines indicate the lattice box and the position of atoms before and after deformation.

Governing equations are obtained from the principle of virtual work: $\delta \mathcal{F} - \delta W = 0$:

$$\sigma_{ij,j}(\mathbf{x}) - \tau_{ijk,jk}(\mathbf{x}) = 0, \quad B_{i,i}(\mathbf{x}) = 0 \quad (7)$$

Let's consider a boundary value problem for the rectangle domain $L \times d$, where $L$ is the length of the film fragment under consideration and $d$ is the thickness of the 2D materials layer. The displacements and potential of magnetic field are assumed as:

$$u_1 = 0, \quad u_3 = a_1 x^2 + a_2 z^2, \quad \psi = b_2 z^2 \quad (8)$$

where $a_1$ and $a_2$ are coefficients corresponding to the strain gradient. Here the coordinate system $(x, z)$ corresponds to $(x_1, x_3)$.

It can be shown from the governing eq. (7) (for details, see Supplementary) that the strain gradient coefficients and for magnetic induction are related with the following equations:

$$\frac{a_1}{a_2} = -\frac{c_{33}}{c_{44}}, \quad b_2 = 0 \quad (9)$$

$$B_3 = 2(\xi_1 + 2\xi_2)a_2 + 2\xi_2 a_1 = 2a_2\left[\xi_1 + \left(2 - \frac{c_{33}}{c_{44}}\right)\xi_2\right] \quad (10)$$

Specification of coefficients by (9) guarantees not only satisfaction of continuum theory governing equations, but results also into the complete set of boundary conditions in the classical continuum theorey for problem A (see Supplementary).

In order to obtain the second equation for unknown coefficients $\xi_1$ and $\xi_2$ another boundary problem should be considered (Fig. 2(b)):

$$u_1 = a_3 x^2 + a_4 z^2, \quad u_3 = 0, \quad \psi = 0 \quad (11)$$

From the governing equations we get relations for strain gradients:

$$a_3 = -\frac{c_{44}}{c_{11}} a_4 \quad (12)$$

and the magnetic induction

$$B_1 = (\xi_1 + 2\xi_2)2a_3 + \xi_2 2a_4 = 2a_4\left[\xi_2 - \frac{c_{44}}{c_{11}}(\xi_1 + 2\xi_2)\right] \quad (13)$$

Finally, we have two expressions for magnetic inductions (10) and (13) with two unknown flexomagnetic coefficients $\xi_1$ and $\xi_2$. If both values $B_1$ and $B_3$ are obtained from DFT calculations, it is easy to get both unknown flexomagnetic parameters.

*Computational details*--Density functional theory simulations were performed within the generalized gradient approximation (GGA) [21] in the form proposed by Perdew-Burke-Ernzerhof (PBE), as implemented in the Vienna Ab-initio Simulation Package (VASP) [22]. The projector augmented wave (PAW) pseudopotentials [23,24] are used. For all the calculations, we chose the energy cutoff to be 500 eV, and an additional effective Hubbard $U_{eff} = 3$ eV for Cr-3$d$ orbitals to deal with the self-interaction error [25], the convergence criterion of the total energy was set to less than $10^{-6}$ eV. We choose the high temperature phase stacking structure, and transform the unit cell into the red line shown in Fig. 1(a) in the bilayer system, optimized to have the lattice parameters of $a = 11.97$ Å and $b = 6.91$ Å. To simulate the experimental condition, we used a nanoribbon composed of 8×1×1 supercell while adding a vacuum larger than 15 Å in the $b$-direction. Thus, the final structure dimensions are $a = 119.78$ Å, $b = 6.91$ Å and $c = 41.39$ Å with $\alpha = \beta = \gamma = 90°$. When the bilayer inside has no strain gradient, the thickness of monolayer and the vdW gap are 3.2 Å and 3.5 Å, respectively, which gives 2 × 3.2 + 3.5 = 9.9 Å thickness of the bilayer. The system contains 64 Cr and 192 I atoms. The 1×5×1 Γ-centered k-grid samplings [26] were adopted for the system. For the mechanical properties, we used the energy-strain method to calculate the elastic constants, generating input files based on VASPKIT [27] with strains ranging from -1% to 1% and fitting the energy to obtain the elastic constants of the system.

*Results and discussion*--As analyzed in the model, we move the atoms in the supercell to simulate three situations, again, the displacements are: $u_1 = 0$, $u_3 = a_1\left(x^2 - z^2 c_{44}/c_{33}\right)$ for boundary condition A while $u_1 = a_3\left(x^2 - a_4 z^2 c_{11}/c_{44}\right)$, $u_3 = 0$ for boundary condition B. The bend deformation means that both boundary condition A and B exist, and note that $a_1 = a_3$ for bend deformation. Fig. 3(b) shows the structures for the three conditions. In order to make the structural features more obvious, the displacement distance of atoms is exaggerated in the figure, and the negative value of strain gradient means that the surface of the bilayer is concave.

First, we calculate the elastic constant of the boundary condition A and B. As described in the Supplementary, we already have the representation of the elastic constant matrix of the orthotropic material, these values in the matrix can be obtained from the fit of the direction-specific strain-energy curve. The energy-strain curves of A and B are shown in Fig. S1, gives the ratio for the elastic constant: $\frac{c_{33}}{c_{44}} \approx 1.80$ and $\frac{c_{44}}{c_{11}} \approx 0.42$. The dependence of strain gradient-induced magnetic moments of CrI$_3$ bilayer per formula unit corresponding to the eq. (10) and (13) are shown in Fig. 3(a). Note that the magnetic moment of the end Cr atoms increases substantially due to the formation of dangling bonds by the unpaired electrons (it's clear to find the Cr's position in Fig. 1(b)). For a more accurate description of the total magnetic moment, we do not consider the contribution of dangling bonds in the total magnetic moment, and as an example we show the effect of dangling bonds in Fig. S2.

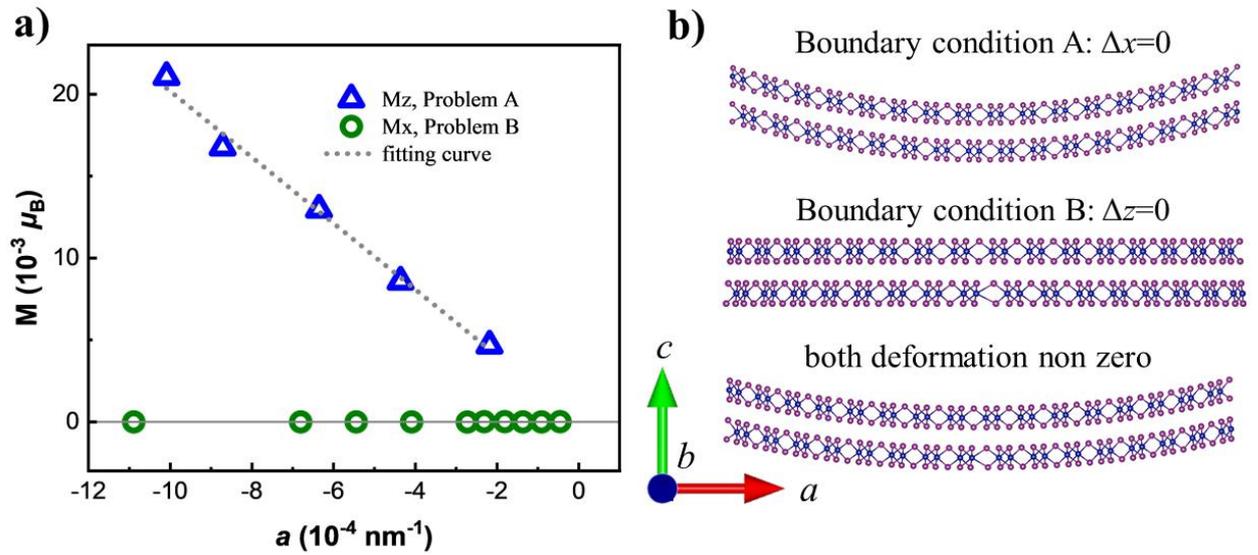

Fig. 3. The calculated flexomagnetic effect: a) the magnetization dependence on the strain gradient parameter along x-axis for boundary conditions A and B. b) The configuration for the boundary conditions A and B, as well as the bending deformation (the superposition of A and B).

Taking into account the data of Fig. 3 one can see that the the flexomagnetic magnetization along x-axis (the Boundary problem B) is negligible with respect to the one along the normal to the plane and is below the accuracy of calculation technique ~ 0.2 $\mu_B$. This result agrees with the Curie principle: the symmetry of the crystal structure (Fig. 1) and symmetry of the "cause" (the deformation, Fig. 2(b)) do not single out any preferential direction in the plane of the bilayer.

Taking into account that $B_1$ is negligibly small the eq. (13) one can estimate the ratio of flexomagnetic constants $\xi_2/\xi_1 \approx 4.5$. The values of flexomagnetic coefficients in accordance to eq. (10) are $\xi_2 \approx 239$ $\mu_0\mu_B\cdot$Å and $\xi_1 \approx 53$ $\mu_0\mu_B\cdot$Å.

Substituting these values of the flexomagnetic coefficients into (eq. (10)) we obtain that the strain gradient-induced magnetic moment along the normal to the plane is proportional to strain gradient $a_2$ with the coefficient 200 $\mu_B\cdot$Å that is about two orders of value larger than the analogous flexomagnetic effect in $Mn_3GaN$ [10].

To rationalize the obtained numerical results let us consider the van der Waals bilayer $CrI_3$ as a system of two oppositely magnetized layers playing the role of sublattices in a conventional antiferromagnet (Fig. 4). From the general arguments the flexomagnetic effect is proportional to the magnetic moment *M* of a single ion in an antiferromagnetic sublattice and to the distance *d* between the pair of antiferromagnetically ordered ions. This distance in the van-der Waals structure of $CrI_3$ bilayer is unusually large for antiferromagnet. The larger is the space separation between antiferromagnetically coupled ions, the more pronounced is the difference in their crystalline environments in the presence of strain gradient. This strain-induced difference in the sublattices' crystal structure results in the imbalance of their magnetizations, *i.e.*, flexomagnetic effect.

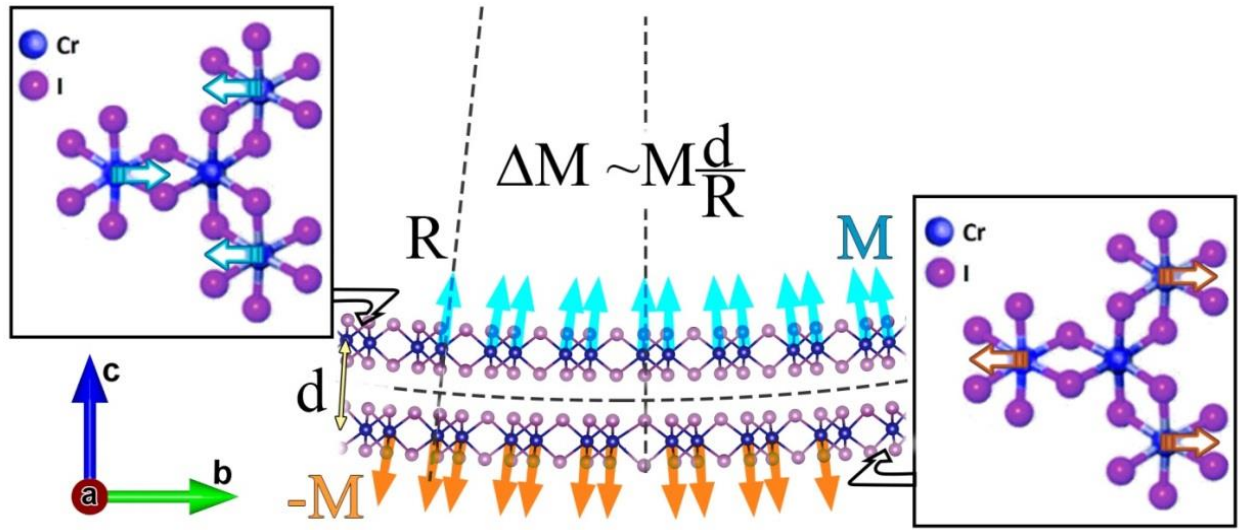

Fig. 4. On the mechanisms of flexomagnetic effect enhancement: **M** is a moment of a single ion in the antiferromagnetic sublattice, $d$ is the distance between the middle lines of the top and bottom CrI$_3$ layers, R is the curvature radius that is inversely proportional to a strain gradient $a_2$ along z-axis. The top views of relative displacements of Cr ions in the top and the bottom layers are shown in the insets

When comparing CrI$_3$ bilayer with the referent flexomagnetic material Mn$_3$GaN one should consider that the magnetic moment of Cr ion M~3 $\mu_B$ is somewhat bigger than M~2 $\mu_B$ for Mn ion and that the antiferromagnetically coupled Cr ion are separated by the spacing $d = 7$ Å while in antiperovskite unit cell of Mn$_3$GaN the distance between Mn ions belonging to different antiferromagnetic sublattices do not exceed 2 Å. However, these factors alone cannot explain two orders of magnitude increase of flexomagnetic effect in CrI$_3$ bilayer compared to Mn$_3$GaN.

Besides these purely geometrical arguments there are also the physical mechanisms of flexomagnetic effect enhancement: the strong dependence of exchange interaction on the distance between atoms (the vivid illustration is RKKY interaction where even the sign of exchange integral changes with distance or the strain-induced modulation of Neel temperature [12]. Within the limits of our model the most probable reason for flexomagnetic effect enhancement is the strain-induced exchange modulation: in the top layer the ferromagnetically ordered Cr ions are closer to each other than in the relaxed state while in the bottom layer they move apart (Fig. 4 insets). The value of intralayer exchange modulation induced by the strain can be estimated from [28]: in the linear approximation the tensile/compressive strain 0.1% corresponds to the reduction/increase of the exchange coupling by 0.5% (see Supplementary Fig.

S3). This results in an effective exchange field for Cr ions in the compressed top layer higher than in the stretched bottom layer resulting in the decompensation of the sublattice magnetizations.

The value of flexomagnetic effect 200 $\mu_B \cdot$Å enables to detect the flexomagnetic effect by high sensitive single spin magnetometry based on a nitrogen–vacancy center (NV-center) [29] if the curvature radius of the ripple is below 1000 Å (that is equivalent to the strain gradient higher than $10^{-3}$ Å$^{-1}$ and strain-induced moment per formula unit 0.2 $\mu_B$)). For the nanotubes of $CrI_3$ similar to those considered in [14] the magnetic moment in the radial direction of the nanotube is an order of value higher, exceeding 1 $\mu_B$ per formula unit.

*Conclusion*--Summarizing, the flexomagnetic effect in $CrI_3$ bilayer is the result of strain-gradient-induced decompensation of antiferromagnetic sublattices and manifest itself on the rippled surface. As soon as the curvature radius of a ripple scales down to the range of hundreds of nanometers and below, the magnetic moment difference per formula unit reaches 1/10 $\mu_B$ that is within the range of single spin NV-center magnetometry. For nanotubes with the radius of 10 nm and below the magnetic moment in radial direction exceeds 1 $\mu_B$, readily detectable by means of magnetic force microscopy. The relatively large value of flexomagnetic effect in the bilayer $CrI_3$ is partly attributable to the large distance between the antiferromagnetically-coupled ions in the van der Waals structure compared to the conventional antiferromagnets, but this geometrical factor alone cannot explain the 2-order enhancement of flexomagnetic coefficient compared to antiperovskite $Mn_3GaN$. To gain insight into the origin of the enhancement the strain-induced ferromagnetic exchange modulation in each layer should be involved. The flexomagnetic effect provides a powerful knob to control magnetic properties of antiferromagnetically coupled van der Waals structures and is interesting for application in spintronics of 2D magnets and straintronics in particular.

This work was supported by the National Natural Science Foundation of China (12074241, 11929401, 52130204), and the Russian Ministry of Science and Education (Project No. 075-15-2022-1131). The support of the Science and Technology Commission of Shanghai Municipality (22XD1400900, 20501130600, 21JC1402600, 21JC1402700), High Performance Computing Center, Shanghai University, and Key Research Project of Zhejiang Laboratory (No. 2021PE0AC02) are acknowledged. L.Q. acknowledges the support of China Scholarship Council. Jan and Vladimir Sladek acknowledge the support by the Slovak Science and Technology Assistance Agency registered under number APVV-18-0004, and the support by the Ministry of Education, Science, Research and Sport of the Slovak Republic under grant number VEGA-2/0061/20. Alexander Pyatakov and Alexey Kaminskiy acknowledge the support of the Basis Foundation "Junior leader" program.